\documentclass[11pt]{article}

\usepackage{fullpage}

\usepackage{url}
\usepackage{amsmath}
\usepackage{graphicx}
\DeclareGraphicsExtensions{.eps,.pdf,.jpeg,.png}
\usepackage{stfloats}
\usepackage{algorithmic}
\usepackage{algorithm}
\usepackage{longtable}

\title{Transparent Checkpoint-Restart over InfiniBand}

\author{Jiajun Cao,\thanks{This work was partially supported
          by the National Science Foundation under Grants~OCI-0960978
	and OCI~1229059.}
   \hbox{\ \ \ \ } Gregory Kerr\footnotemark[1],
   \hbox{\ \ \ \ } Kapil Arya\footnotemark[1],
   \hbox{\ \ \ \ } Gene Cooperman\footnotemark[1] \\
College of Computer and Information Science\\
Northeastern University\\
Boston, MA 02115 / USA\\
Email: jiajun@ccs.neu.edu, kerrgi@gmail.com, \{kapil,gene\}@ccs.neu.edu
}

\date{}

\begin{document} 

\maketitle

\begin{abstract}
InfiniBand is widely used for low-latency, high-throughput cluster
computing. Saving the state of the InfiniBand network as part of
distributed checkpointing has been a long-standing challenge for
researchers. Because of a lack of a solution, typical MPI implementations
have included custom checkpoint-restart services that ``tear down''
the network, checkpoint each node as if the node were a standalone
computer, and then re-connect the network again. We present the first
example of transparent, system-initiated checkpoint-restart that
directly supports InfiniBand.  The new approach is independent of
any particular Linux kernel, thus simplifying the current practice
of using a kernel-based module, such as BLCR.  This direct approach
results in checkpoints that are found to be faster than with the use
of a checkpoint-restart service.  The generality of this approach is
shown not only by checkpointing an MPI computation, but also a native
UPC computation (Berkeley Unified Parallel~C), which does not use MPI.
Scalability is shown by checkpointing 2,048~MPI processes across
128~nodes (with 16~cores per node).  In addition, a cost-effective
debugging approach is also enabled, in which a checkpoint image from an
InfiniBand-based production cluster is copied to a local Ethernet-based
cluster, where it can be restarted and an interactive debugger can be
attached to it.  This work is based on a plugin that extends the DMTCP
(Distributed MultiThreaded CheckPointing) checkpoint-restart package.
\end{abstract}

\section{Introduction}
\label{sec:Introduction}

InfiniBand is the preferred network for most of high performance
computing and for certain Cloud applications, due to its low latency.
We present a new approach to checkpointing over InfiniBand.
This is the first efficient and transparent solution for {\em direct}
checkpoint-restart over the InfiniBand network (without the intermediary
of an MPI implementation-specific checkpoint-restart service).  This also
extends to other language implementations over InfiniBand, such as
Unified Parallel~C (UPC).  This work contains several subtleties,
such as whether to drain the ``in-flight data'' in an InfiniBand
network prior to checkpoint, or whether to force a re-send of data
after resume and restart.  A particularly efficient solution was found,
which combines the two approaches.  The InfiniBand completion queue is
drained and refilled on resume or restart, and yet data is never re-sent,
{\em except} in the case of restart.

Historically, transparent (system-initiated) checkpoint-restart
has typically been the first technology that one examines in order
to provide fault tolerance during long-running computations.
For example, transparency implies that a checkpointing scheme
works independently of the programming language being used (e.g.,
Fortran, C, or C++).  In the case of distributed computations over
Ethernet, several distributed checkpointing approaches have been
proposed~\cite{CoopermanAnselMa06,Cruz05,LaadanEtAl05,LaadanEtAl07,Chpox}.
Unfortunately, those solutions do not extend to supporting the InfiniBand
network.  Other solutions for distributed checkpointing are specific to a
particular MPI implementation~\cite{Bouteiler06,GaoEtAl06,HurseyEtAl09,
OpenMPICheckpoint07,KerrEtAl11,LemarinierEtAl04,
CheckpointLAMMPI05b,SankaranEtAl05}.  These MPI-based checkpoint-restart
services ``tear down'' the InfiniBand connection, after which a
single-process checkpoint-restart package can be applied.

Finally, checkpoint-restart is the process of saving to stable storage (such
as disk or SSD) the state of the processes in a running computation,
and later re-starting from stable storage.  {\em Checkpoint} refers
to saving the state, {\em resume} refers to the original process
resuming computation, and {\em restart} refers to launching a new
process that will restart from stable storage.  The checkpoint-restart is
{\em transparent} if no modification of the application is required.
This is sometimes called {\em system-initiated} checkpointing.

Prior schemes specific to each MPI implementation had: (i)~blocked the
sending of new messages and waited until pending messages have been
received; (ii)~``torn down'' the network connection; (iii)~checkpointed
each process in isolation (without the network), typically using the
BLCR package~\cite{BLCR03,BLCR06} based on a kernel module; (iv)~and
then re-builds the network connection.  Those methods have three drawbacks:
\begin{enumerate}
\item Checkpoint-resume can be slower due to the need to tear down and
	re-connect the network.
\item PGAS languages such as UPC~\cite{UPC02} must be re-factored to
	run over MPI instead
	of directly over InfiniBand in order to gain support for
	transparent checkpoint-restart.  This can produce additional
	overhead.
\item The use of a BLCR kernel module implied that the restart cluster
	must use the same Linux kernel as the checkpoint cluster.
\end{enumerate}

The current work is implemented as a plugin on top of DMTCP (Distributed
MultiThreaded CheckPointing)~\cite{AnselEtAl09}.  The experimental
evaluation demonstrates DMTCP-based checkpointing of Open~MPI for
the NAS LU benchmark and others.  For 512~processes, checkpointing
to a local disk drive, occurs in 232~seconds, whereas it requires
36~seconds when checkpointing to back-end Lustre-based storage.
Checkpointing of up to 2,048 MPI processes (128~nodes with 16~cores
per node) is shown to have a run-time overhead between 0.8\% and 1.7\%.
This overhead is shown to be a little less than the overhead when using
the checkpoint-restart of Open~MPI using BLCR.  Tests were
also carried out on Berkeley UPC~\cite{UPC99} over GASNet's ibv
conduit~\cite{GASNet02}, with similar results for checkpoint times and
run-time overhead.

The new approach can also be extended to provide capabilities
similar to the existing interconnection-agnostic use of the Open~MPI
checkpoint-restart service~\cite{HurseyEtAl09}.  Specifically, we
demonstrate an additional IB2TCP plugin that supports cost-effective use
of a symbolic debugger (e.g., GDB) on a large production computation.
The IB2TCP plugin enables checkpointing over InfiniBand and restarting
over Ethernet.  Thus, when a computation fails after on a production
cluster (perhaps after hours or days), the checkpoint image can be
copied to an inexpensive, smaller, Ethernet-based cluster for interactive
debugging.  An important contribution of the IB2TCP plugin, is that unlike
the BLCR kernel-based approach, the DMTCP/IB2TCP approach supports using
an Ethernet-based cluster that uses a different Linux kernel, something
that occurs frequently in practice.

Finally, the approach of using DMTCP plugins provides is easily
maintainable, as measured by lines of code.  The primary plugin,
supporting checkpointing over InfiniBand, consists of 2,700~lines of code.
The additional IB2TCP plugin consists of 1,000~lines of code.

\paragraph{Organization of Paper.}
The rest of this paper includes
the background of DMTCP (Section~\ref{sec:dmtcp}. and
the algorithm for checkpointing over InfiniBand (Section~\ref{sec:algorithm}).
Limitations of this approach are discussed in Section~\ref{sec:limitations}.
An experimental evaluation is presented in Section~\ref{sec:experiments}.
Section~\ref{sec:ib2tcp} is of particular note, in reporting on the
IB2TCP plugin for migrating a computation from a production cluster
to a debug cluster.
Finally, the related work (Section~\ref{sec:relatedWork})
and conclusions (Section~\ref{sec:conclusion}) are presented.

\section{Background: InfiniBand and DMTCP}
\label{sec:dmtcp}

Section~\ref{sec:infiniband} reviews some concepts of InfiniBand,
necessary for understanding the checkpointing approach described
in Section~\ref{sec:algorithm}.
Section~\ref{sec:plugins} describes the use of plugins in DMTCP.

\subsection{Review of InfiniBand ibverbs API}
\label{sec:infiniband}

\begin{figure}[hbt]
\begin{center}
\includegraphics[width=0.9\columnwidth]{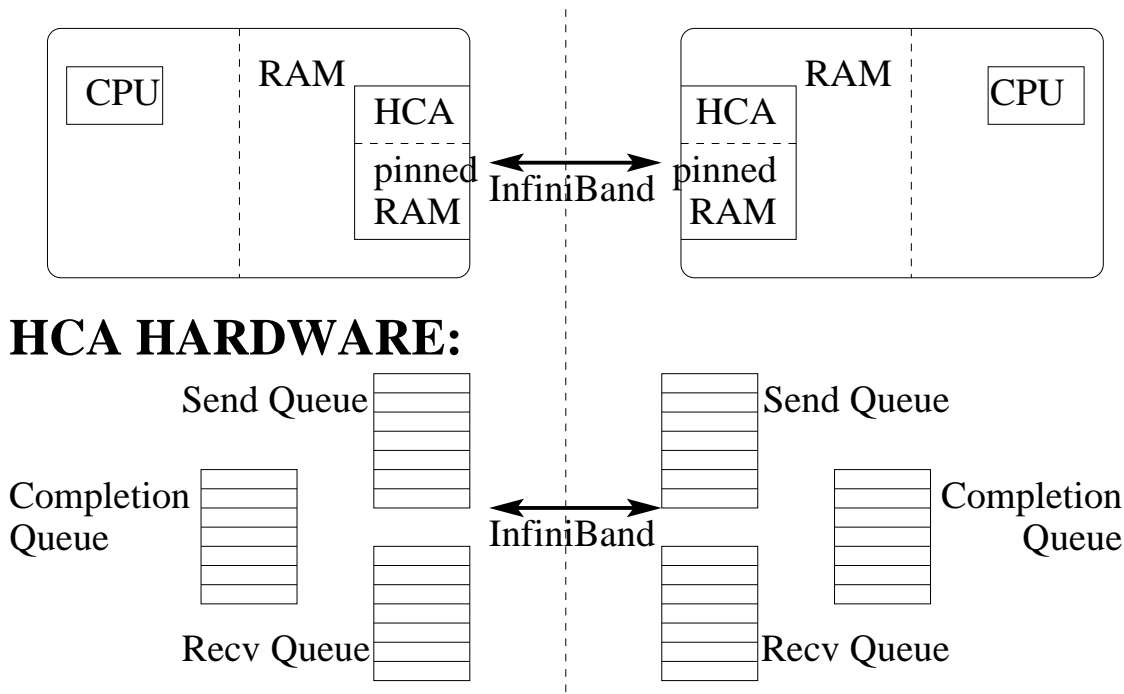}
\end{center}
\caption{\label{fig:ibConcepts} InfiniBand Concepts}
\end{figure}

In order to understand the algorithm, we review some concepts
from the Verbs API of InfiniBand.  While there are several references
that describe InfiniBand, we recommend one of~\cite{KerrIB11,BedeirIB10}
as a gentle introduction for a general audience.

Recall that the InfiniBand network uses {\em RDMA} (remote DMA to the RAM
of a remote computer).  Each computer node must have a Host Channel
Adapter (HCA) board
with access to the system bus (memory bus).
With only two computer nodes, the HCA adapter boards may be connected
directly to each other.  With three or more nodes, communication must
go through an InfiniBand switch in the middle.  Note also that the
bytes of an InfiniBand message may be delivered out of order.

Figure~\ref{fig:ibConcepts} reviews the basic elements of an InfiniBand
network.  A hardware host channel adapter (HCA) and the software library
and driver together maintain at least one {\em queue pair} and a {\em
completion queue} on each node.  The queue pair consists of a send queue
on one node and a receive queue on a second node.  In a bidirectional
connection, each node will contain both a send queue and a receive queue.
Sending a message across a queue pair causes an entry to be added to
the completion queue on each node.  However, it is possible to set a
flag when posting a work request to the send queue, such that no entry
is added to the completion queue on the ``send'' side of the connection.

Although not explicitly introduced as a standard, libibverbs
(provided by the Linux OFED distribution) is the most
commonly used InfiniBand interface library.   We will describe
the model in terms of the functions prefixed by {\tt ibv\_} for
the {\em verbs library} (libibverbs).  Many programs also
use OFED's convenience functions, prefixed by {\tt rdma\_*}.
OFED also provides an optional library, librdmacm (RDMA connection
manager) for ease of connection set-up and tear-down in conjunction
with the verbs interface.  Since this applies only to set-up and tear-down,
this library does not affect the ability to perform
transparent checkpoint-restart.

We assume the reliable connection model (end-to-end context),
which is by far the most commonly used model for InfiniBand.
There are two models for the communication:
\begin{itemize}
\item Send-receive model
\item RDMA (remote DMA) model (often employed for efficiency,
	and serving as the inspiration for the one-sided
	 communication of the MPI-2 standard)
\end{itemize}
Our InfiniBand plugin supports both models, and a typical MPI implementation
can be configured to use either model.

\subsubsection{Send-Receive model of InfiniBand communication}
\label{sec:sendReceiveModel}

We first describe the steps in processing the send-receive
model for InfiniBand connection.  It may be useful to examine
Figure~\ref{fig:ibConcepts} while reading the steps below.
\begin{enumerate}
 \item Initialize a hardware context, which causes a buffer in RAM
	to be allocated.  All further operations are with respect to
	this hardware context.
 \item Create a protection domain that sets the permissions to determine
	which computers may connect to it.
 \item Register a memory region, which causes the virtual memory to
	be pinned to a physical address (so that the operating system
	will not page that memory in or out).
 \item A completion queue is created for each of the sender and
	the receiver.  This completion queue will be used later.
 \item Create a queue pair (a send queue and a receive queue) associated
	with the completion queue.
 \item An end-to-end connection is created between two queue pairs, with
	each queue pair associated with a port on an HCA adapter.
	The sender and receiver queue pair information (several ids)
	is exchanged, typically either using either TCP
	(through a side channel that is not non-InfiniBand),
	or by using an rdmacm library whose API is transport-neutral.
 \item The receiver creates a work request and posts it to the receive queue.
	(One can post multiple receive buffers in advance.)
 \item The sender creates one or more work requests and posts them
	to the send queue.
 \item
\label{step:postReceive} {\em NOTE:\/} The application must ensure that
	a receive buffer has been posted before it posts a work request
	to the send queue.  It ia an application error if this is not
	the case.
 \item The transfer of data now takes place between a posted buffer on
	the send queue and a posted buffer on the receive queue.
	The posted send and receive buffers have now been used up,
	and further posts are required for further messages.
 \item
\label{step:completion}
	Upon completion, work completions are generated by the hardware
	and appended to each of the completion queues, one queue on the
	sender's node and one queue on the receiver's node.
 \item
\label{step:polling}
        The sender and receiver each poll the completion queue until a
	work completion is encountered.  (A blocking request for
	work completion also exists as an alternative.  A blocking
	request must be acknowledged on success.)
 \item Polling causes the work completion to be removed from the
	completion queue.  Hence, further polling will eventually see
	further completion events.  Both blocking and non-blocking versions
	of the polling calls exist.
\end{enumerate}

We also remark that a work request (a WQE or Work Queue Entry) points
to a list of scatter/gather elements, so that the data of the message
need not be contiguous.

\subsubsection{RDMA model of InfiniBand communication}
The RDMA model is similar to the send-receive model.  However, in this
case, one does not post receive buffers.  The data is received directly
in a memory region.  An efficient implementation of MPI's one-sided
communication (MPI\_Put, MPI\_Get, MPI\_Accumulate), when implemented
over InfiniBand, will typically employ the RDMA model~\cite{JiangEtAl04}.

As a consequence, Step~\ref{step:postReceive} of
Section~\ref{sec:sendReceiveModel} does not appear in the RDMA model.
Similarly, Steps~\ref{step:completion} and~\ref{step:polling} are modified
in the RDMA model to refer to completion and polling solely for the send
end of the end-to-end connection.

Other variations exist, which are supported in our work, but not
explicitly discussed here.  In one example, an InfiniBand application may
choose to send several messages without requesting a work completion in
the completion queue.  In these cases, an application-specific algorithm
will follow this sequence with a message that includes a work completion.
In a second example, an RDMA-based work request may request an immediate
mode, in which the work completion is placed only in the remote completion
queue and not in the local completion queue.

\subsection{DMTCP and Plugins}
\label{sec:plugins}

DMTCP is a transparent, checkpoint-restart package that supports
third-party plugins.  The  current work on InfiniBand support was
implemented as a DMTCP plugin~\cite{dmtcpPlugin}.  The plugin is used
here to virtualize the InfiniBand resources exposed to the end user,
such as the queue pair struct (ibv\_qp) (see Figure~\ref{fig:ibConcepts}).
This is needed since upon restart from a checkpoint image, the plugin
will need to create a new queue pair for communication.  As a result,
the InfiniBand driver will create a new queue pair struct at a new
address in user space, with new ids.

Plugins provide three core features used here to support virtualization:
\begin{enumerate}
  \item wrapper functions around functions of the InfiniBand library:
	these wrappers translate between virtual resources (see by the target
	application) and real resources (seen within the InfiniBand library,
	driver and hardware).  The wrapper function also records changes to
	the queue pair and other resources for later replay during restart.
  \item event hooks:  these hooks are functions within the plugin that DMTCP
	will call at the time of checkpoint and restart.  Hence, the plugin
	is notified at the time of checkpoint and restart,  so as to
	update the virtual-to-real translations, to recreate the network
	connection upon restarting from a checkpoint image, and to replay
	some information from the logs.
  \item\label{step:publishSubscribe}a publish/subscribe facility:
        to exchange ids
	among plugins running on the different computer nodes whenever
	new device connections are created.  Examples of such ids are
	local and remote queue pair numbers and remote keys of memory regions.
\end{enumerate}

\section{Algorithm}
\label{sec:algorithm}

\begin{figure}[hbt]
\begin{center}
\includegraphics[width=0.9\columnwidth]{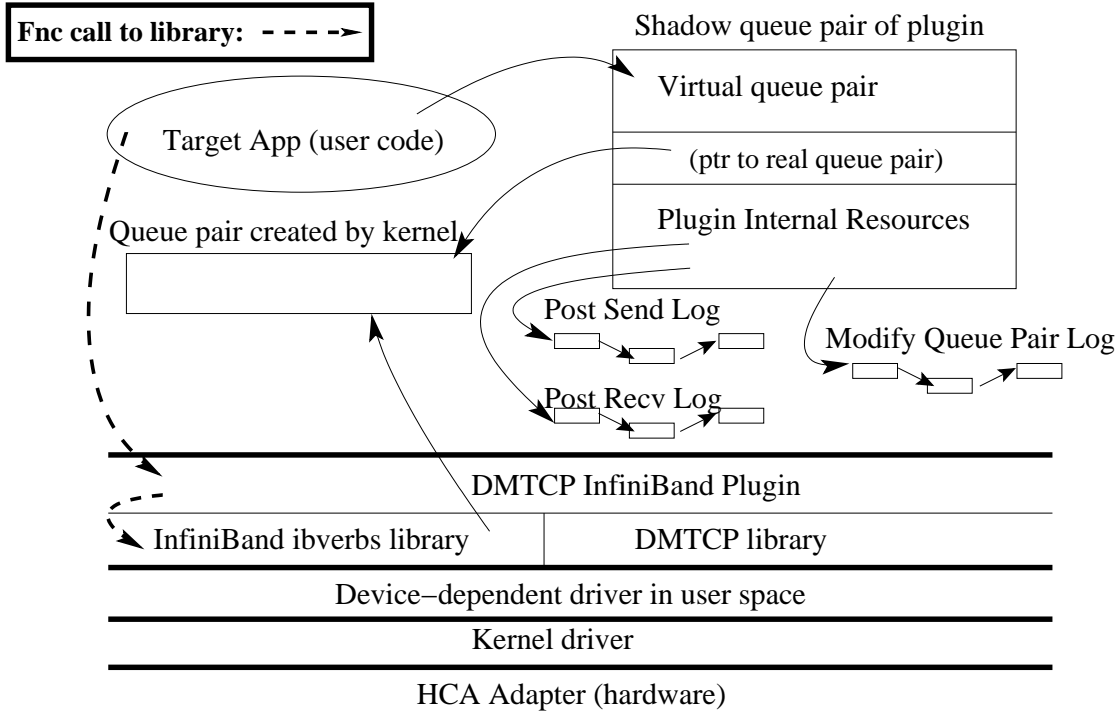}
\end{center}
\caption{\label{fig:qp} Queue pair resources and their virtualization.
   (The plugin keeps a log of calls to post to or to modify the queue pair.)}
\end{figure}

Figure~\ref{fig:qp} presents an overview of the virtualization of a
queue pair.  This is the most complex of the subsystems being checkpointed.
In overview, observe that the DMTCP plugin library interposes between
most calls from the target application to the InfiniBand ibverbs library.
This allows the DMTCP InfiniBand plugin to intercept the creation of
a queue pair by the InfiniBand kernel driver, and to create a shadow
queue pair.  The target application is passed a pointer only to the
virtual queue pair created by the plugin.  Thus, any further ibverbs calls
to manipulate the queue pair will be intercepted by the plugin, and
appropriate fields in the queue pair structure can be appropriately
virtualized before the real ibverbs call.

Similarly, any ibverbs calls to post to the send or receive queue or
to modify the queue pair are intercepted and saved in a log.  This log
is used for internal bookkeeping by the plugin, to appropriately model
work requests as they evolve into the completion queue.  In particular,
note that a call to ibv\_post\_send may request that no work completion
be entered on the completion queue.

\subsection{The Checkpoint-Restart Algorithm}

As the user base code makes calls to the verbs library, we will use DMTCP
plugin wrapper functions around these library functions to interpose.
Hence the user call goes first to our DMTCP plugin library.  We then
extract parameters describing how the resources were created, before
passing on the call to the verbs library, and later passing back the
return value.  This allows us to recreate semantically equivalent
copies of those same resources on restart {\em even if we restart on
a new computer}.

in particular, we record any calls to {\tt modify\_qp} and to
{\tt modify\_srq}.  On restart, those calls are replayed in order
to place the corresponding data structures in a semantically equivalent
state to pre-checkpoint.

While the description above appears simple, several subtleties arise,
encapsulated in the following principles.

\paragraph{Principle 1:  Never let the user see a pointer to the actual
	 InfiniBand resource.}

A verbs call that creates a new InfiniBand resource
will typically create a struct, and return a pointer to that struct.
We will call this struct created by the verbs library a {\em real struct}.
If the end user code create an InfiniBand resource,
we interpose to copy that struct to a a new {\em shadow struct},
and then pass back to the end user the pointer to this shadow struct.
Some examples of InfiniBand resources for which this is done are:
a context, a protection domain, a memory region, and a queue pair.

The reason for this is that many implementations of InfiniBand libraries
contain additional undocumented fields in these structs, in addition to
those documented by the corresponding ``man page''.  When we restart
after checkpoint, we cannot pass the original pre-checkpoint struct
to the verbs library.  The undocumented (hidden) fields would not match
the current state of the InfiniBand hardware on restart.  (New device-dependent
ids will be in use after restart.)

So, on restart, we create an entirely new InfiniBand
resource (using the same parameters as the original).  This new struct
should
be semantically equivalent to the pre-checkpoint original, and the
hidden fields will correspond to the post-restart state of the hardware.

One can think of this as a form of virtualization.  The user is
passed a pointer to a {\em virtual struct}, the shadow struct.
The verbs library knows only about the {\em real struct}.
So, we will guarantee that the verbs library only sees real structs,
and that the end user code will only see virtual structs.

To do this, we interpose
our DMTCP plugin library function if a verbs library function refers
to one of these structs representing InfiniBand resources.  If the end
user calls a verbs library function that
returns a pointer to a real struct, then our interposition will replace
this and return a pointer to a corresponding
virtual struct.  If the user code passes an argument pointing to a virtual
struct, we will replace it by a pointer to a real struct before calling
the verbs library function.

\paragraph{Principle 2:  Inline functions almost always operate through
 pointers to possibly device-dependent functions.  Add wrappers around
 the pointers, and not the inline functions.}

The previous principle depends on wrapper functions that interpose in order
to translate between real and virtual structs.  But in the OFED ibverbs
implementation,
some of the apparent library calls to the verbs library are in fact
inline functions.  A DMTCP plugin cannot easily interpose on inline functions.
Luckily, these inline functions are often associated with possibly
device-dependent functions.  So, the OFED software design expands the
inline functions into calls to pointers to other functions.
Those pointers can be found through the {\tt ops} field of a
{\tt struct ibv\_context}.
The {\tt ops} field is of type {\tt struct ibv\_context\_ops} and contains
the function pointers.  This gives us access,
and we modify the function pointers to point to our own functions,
which can then interpose and finally call the original function pointers
created by the verbs library.

\paragraph{Principle 3:  Carry out bookkeeping on posts of work queue entries
 to the send and receive queue.}

As work requests are entered onto a send queue or receive queue, the
wrapper functions of the DMTCP plugin record those work requests (which
have now become work queue entries).  When the completion queue is polled,
if a completion event corresponding to that work queue entry is found,
then the DMTCP plugin records that the entry has been destroyed.  At the
time of checkpoint, there is a log of those work queue entries that have
been posted and not yet destroyed.  At the time of restart, the send and
receive queues will initially be empty.  So, those work queue entries
are re-posted to their respective queues.  (In the case of resume,
the send and receive queues continue to hold their work queue entries,
and so no special action is necessary.)

\paragraph{Principle 4:  At the time of checkpoint, ``drain'' the completion
  queue of its completion events.}

At the time of checkpoint, and after all user threads have been quiesced,
the checkpoint thread polls the completion queue for remaining completion
events not yet seen by the end user code.  A copy of each completion
event seen is saved by the DMTCP plugin.  Note that we must drain the
completion queue for each of the sender and the receiver.  Recall also
that the verbs library function for polling the completion queue will
also remove the polled completion event from the completion queue as it
passes that event to the caller.

\paragraph{Principle 5:  At the time of restart or resume, ``refill''
  a virtual completion queue.}

At the time of restart or resume and before any user threads have
been re-activated, we must somehow refill the completion queue, since
the end user has not yet seen the completion events from the previous
principle.  To do this, the DMTCP plugin stores the completion events
of the previous principle in its own private queue.  The DMTCP plugin
library then interposes between any end user calls to a completion queue
and the corresponding verbs library function.  If the end user polls
the completion queue, the DMTCP wrapper function passes back to the end
user the plugin's private copy of the completion events, and the verbs
library function for polling is never called.  Only after the private
completion queue becomes empty are further polling calls passed on to
the verbs library function.  Hence, the plugin's private queue
becomes part of a {\em virtual completion queue}.

\paragraph{Principle 6:  Any InfiniBand messages still ``in flight'' can
 be ignored.}

If data from an InfiniBand message is still in flight (has not yet
arrived in the receive buffer), then InfiniBand will not generate a
completion event.  Note that the InfiniBand hardware may continue
to transport the data of a message, and even generate a completion
event {\em after all user threads have been quiesced for checkpoint}.
Nevertheless, a simple rule operates.

If our checkpoint thread has not seen a completion event that arrived
late, then we will not have polled for that completion event.  Therefore,
our bookkeeping in Principle~3 will not have removed the send or receive
post from our log.  Further, this implies that the memory buffers
will continue to have the complete data, since it was saved on checkpoint
and restored on restart.  Therefore, upon restart (which implies a fresh,
empty completion queue), the checkpoint thread will issue another
send or receive post (again following the logic of Principle~3).

\paragraph{Remark.}
Blocking requests for a completion event ({\tt ibv\_get\_cq\_event}) and for
shared receive queues create further issues.
While those details add some complication, their solution is straightforward
and is not covered here.

\subsection{Virtualization of InfiniBand Ids on Restart}
\label{sec:virtualization}

A number of InfiniBand objects and associated ids will change on restart.  All of these must be virtualized.  Among these objects and ids
are
ibv contexts,
protection domains,
memory regions
(the local and
remote keys (lkey/rkey) of the memory regions),
completion queues,
queue pairs
(the queue pair number, qp\_num),
and the local id of the HCA port being used (lid).
Note that the lid of an HCA port will not change if restarting on the
same host, but it may change needed when restarting on a new host,
which may have been configured to use a different port.

In all of the above cases, the plugin assigns a virtual id and maintains
a translation table between virtual and real id.  The application
sees only the virtual~id.  Any InfiniBand calls are processed through
the plugin, where virtual ids are translated back to real~ids.

On restart, the InfiniBand hardware may assign new real~ids for
a given InfiniBand object.  In this case, the real~ids are updated
within the translation tables maintained by the plugin.

\subsubsection{Virtualization of remote ids: rkey, qp\_num and lid}

However, a more difficult issue occurs in the case of remote
memory keys (rkey), queue pair numbers (qp\_num) and local ids (lid).
In all three cases, an InfiniBand application must pass these ids
to a remote node for communication with the local node.  The remote
need will need the qp\_num and lid when calling {\tt ibv\_modify\_qp}
to initialize a queue pair that connects to the local node.
The remote node will need the rkey when calling {\tt ibv\_post\_send}
to send a message to the local node.

Since the plugin allows the application to see only virtual ids, the
application will employ a virtual id when calling {\tt ibv\_modify\_qp}
and {\tt ibv\_post\_send}.  The plugin will first
replace the virtual id by the real~id, which is known to the InfiniBand
hardware.  To do this, the plugin within each remote node must contain
a virtualization table to translate all virtual ids by real ids.

Next, we recall how a remote node received a virtual~id in the first
place.  The InfiniBand specification solves this bootstrapping problem
by requiring the application to pass these three ids to the remote node
through some out-of-band mechanism.  When the application employs this
out-of-band mechanism, the remote node will ``see'' the virtual ids
that the plugin passed back to the application upon completion of an
InfiniBand call.

The solution chosen for the InfiniBand plugin is that assigns a
virtual~id, which is the same as the real~id at the time of the initial
creation of the InfiniBand object.  After restart, the InfiniBand
hardware may change the real~id.  At the time of restart, the plugin
uses the DMTCP coordinator and the publish-subscribe feature to exchange
the new real~ids, associated with a given virtual~id.  Since the application
continues to see only the virtual~ids, the plugin can continue
to translate between virtual and real~ids through any wrapper by
which the application communicates to the InfiniBand hardware (see
Figure~\ref{fig:qp}.  (A subtle issue can arise if a queue pair or memory
region is created after restart.  This is a rare case.  Although we
have not seen this in the current work, Section~\ref{sec:limitations}
discusses two possible solutions.)

\subsubsection{Virtualization of rkeys}
\label{sec:rkey}

Next, the case of rkeys (remote memory region keys) poses a particular
problem that does not occur for queue pair numbers or local ids.
This is because an rkey is guaranteed unique by InfiniBand only with
respect to the protection domain within which it was created.  Thus,
if a single InfiniBand node has received rkeys from many remote nodes,
then the rkeys for two different remote nodes may conflict.

Normally, InfiniBand can resolve this conflict because a queue pair must
be specified in order to send or receive a message.  The local queue
pair number determines a unique queue pair number on the remote node.
The remote queue pair number then uniquely determines an associated
protection domain~$pd$.  With the remote~$pd$, all rkeys are unique.
Hence, the InfiniBand driver on the remote node uses the ($pd$, rkey)
pair, to determine a unique memory address on the remote node.

In the case of the InfiniBand plugin, the vrkey and rkey are identical
if no restart has taken place.  (It is only after restart that the
rkey may change, for a given vrkey).  Hence, prior to the first
checkpoint, translation from vrkey to rkey is trivial.

After a restart, the InfiniBand plugin must employ a strategy motivated
by that of the InfiniBand driver.  In
a call to {\tt ibv\_post\_send}, the target application will provide
both a virtual queue pair number and a virtual rkey (vrkey).
Unlike InfiniBand, the plugin must translate the vrkey into the real
rkey on the local node.  However, during a restart, each node has
published its locally generated rkey, the corresponding~$pd$ (as
a globally unique id; see above), and the corresponding vrkey.
Similarly, each node has published the virtual queue pair number and
corresponding~$pd$ for any queue pair generated on that node.
Each node has also subscribed to the above information published by all
other nodes.

Hence, the local node is aware of the following through publish-subscribe
during restart:
\begin{align*}
  &{\rm (qp\_num, pd)} \\
  &{\rm (vkey, pd, rkey)}
\end{align*}

The call to {\tt ibv\_post\_send} provided the local virtual qp\_num, and
the vrkey.  This allows us to translate into the remote virtual qp\_num,
and hence the remote real qp\_num.  This allows us to derive the globally
unique $pd$ from the tuples above.  The $pd$ and $vkey$ together then
allow us to use the second tuple to derive the necessary rkey, to be
used when calling the InfiniBand hardware.

\section{Limitations}
\label{sec:limitations}

Recall that DMTCP copies and restores all of user-space memory.
In reviewing Figure~\ref{fig:qp}, one notes that the user-space memory
includes a low-level device-dependent driver in user space.  If, for
example, one checkpoints on a cluster partition using Intel/QLogic, and if
one restarts on a Mellanox partition, then the Mellanox low-level driver
will be missing.  This presents a restriction for heterogeneous computing
centers in the current implementation.

In future work, we will consider one of two alternative implementations.
First, it is possible to implement a generic ``stub'' driver library,
which can then dispatch to the appropriate device-dependent library.
Second, it is possible to force the InfiniBand library to re-initialize
itself by restoring the pre-initialization version of the InfiniBand
library data segment, instead of the data segment as it existed just
prior to checkpoint.  This will cause the InfiniBand library to appear
to be uninitialized, and it will re-load the appropriate device-dependent
library.

Another issue is that the InfiniBand hardware may post completions to the
sender and receiver at slightly different times.  Thus, after draining
the completion queue, the plugin waits for a fraction of a second,
and then drains the completion queue one more time.  This is repeated
until no completions are seen in the latest period.  Thus, correctness is
affected only if the InfiniBand hardware posts corresponding completions
relatively far apart in time, which is highly unlikely.
(Note that this situation occurs in two cases: InfiniBand send-receive
mode; and InfiniBand RDMA mode for the special case of ibv\_post\_send
while setting the immediate data flag.)

In a related issue, when using the immediate data flag or the inline
flag in the RDMA model, a completion is posted only on the
receiving node.  These flags are intended primarily for applications
that send small messages.  Hence, the current implementation sleeps
for a small amount of time to ensure that such messages complete.
A future implementation will use the DMTCP coordinator to complete
the bookkeeping concerning messages sent and received, and will
continue to wait if needed.

Next, the current implementation does not support unreliable connections
(the analog of UDP for TCP/IP).  Most target applications do
not use this mode, and so this is not considered a priority.

A small memory leak on the order of hundreds of bytes per restart
can occur because the memory for the queue pair struct and other
data structures generated by by InfiniBand driver are not recovered.
While techniques exist to correct this issue, it is not considered
important, given the small amount of memory.  (Note that this is
not an issue for registered or pinned memory, since on restart,
a fresh process does not have any pinned memory.)

\section{Related Work}
\label{sec:relatedWork}

The implementation described here can be viewed as interposing a
shadow device driver between the end user's code and the true device
driver.  This provides an opportunity to virtualize the fields of the
queue pair struct seen by the end user code.  Thus, the InfiniBand
driver is modelled without the need to understand its internals.
This is analogous the idea of using a shadow kernel device by Swift
\hbox{et~al.}~\cite{SwiftEtAl04,SwiftEtAl06}.  In that work, after a
catastrophic failure by the kernel device driver, the shadow device
driver was able to take over and place the true device driver back in a
sane state.  In a similar manner, restarting on a new host with a new
HCA Adapter can be viewed as a catastrophic failure of the InfiniBand
user-space library.  Our virtual queue pair along with the log of pending
posts and modifications to the queue pair serves as a type of shadow
device driver.  This allows us to place back into a sane state the HCA
hardware, the kernel driver and the device-dependent user-space driver.

This work is based on DMTCP (Distributed MultiThreaded
CheckPointing)~\cite{AnselEtAl09}.  The DMTCP project began in
2004~\cite{CoopermanAnselMa05,CoopermanAnselMa06}.  With the
development of DMTCP versions~2.x, it has emphasized the use of
plugins~\cite{dmtcpPlugin} for more modular maintainable code.

Currently, BLCR~\cite{BLCR06} is widely used as one component of an
MPI dialect-specific checkpoint-restart service.  This design is
fundamentally different, since an MPI-specific checkpoint-restart
service calls BLCR, whereas DMTCP transparently invokes an arbitrary
MPI implementation.  Since BLCR is kernel-based, it provides direct
support only on one computer node.  Most MPI dialects overcome
this in their checkpoint-restart service by disconnecting any
network connections, delegating to BLCR the task of a single-node
checkpoint, and then reconnect the network
connection.  Among the MPI implementations using BLCR are
Open~MPI~\cite{OpenMPICheckpoint07} (CRCP coordination protocol),
LAM/MPI~\cite{CheckpointLAMMPI05b}, MPICH-V~\cite{Bouteiler06}, and
MVAPICH2~\cite{GaoEtAl06}.  Other MPI implementations provide their own
analogs~\cite{GaoEtAl06,LemarinierEtAl04,CheckpointLAMMPI05b,SankaranEtAl05}.
In some cases, an MPI implementation may support an {\em
application-initiated} protocol in combination with BLCR (such
as SELF~\cite{OpenMPICheckpoint07,CheckpointLAMMPI05b}).  For
application-initiated checkpointing, the application writer guarantees
that there are no active messages at the time of calling for a checkpoint.

Some recommended technical reports for further reading on the design
of InfiniBand are~\cite{BedeirIB10,KerrIB11}, along with the earlier
introduction to the C~API~\cite{WoodruffEtAl05}.  The report~\cite{KerrIB11}
was a direct result of the original search for a clean design in
checkpointing over InfiniBand, and \cite{KerrEtAl11} represents
a talk on interim progress.

In addition to DMTCP, there have been several packages for transparent,
distributed checkpoint-restart of applications running over TCP
sockets~\cite{Cruz05,LaadanEtAl07,LaadanEtAl05,Chpox}.  The first two
packages (\cite{Cruz05} and~\cite{LaadanEtAl05,LaadanEtAl07}) are based
on the Zap package~\cite{zap02}.

The Berkeley language Unified Parallel~C (UPC)~\cite{UPC02} is an example
of a PGAS language (Partitioned Global Address Space).  It runs
over GASNet~\cite{GASNet02} and evolved from
experience with earlier languages for DSM (Distributed Shared Memory).

\section{Experimental Evaluation}
\label{sec:experiments}

The experiments are divided into four parts:
scalability with
more nodes in the case of Open~MPI (Section~\ref{sec:scalability});
comparison between BLCR and
DMTCP for MPI-based computations (Section~\ref{sec:comparison});
tests on Unified Parallel~C (UPC) (Section~\ref{sec:upc});
and demonstration of migration from InfiniBand to TCP
(Section~\ref{sec:ib2tcp}).

The LU benchmark from the NAS parallel benchmarks was used except
for tests on UPC, for which there was no available port of the
NAS LU benchmark.

Note that the default behavior of DMTCP is to compress checkpoint images
using gzip.  All of the experiments used the default gzip invocation,
unless otherwise noted.  As shown in Section~\ref{sec:scalability},
gzip produced almost no additional compression, but it is also true
that the additional CPU time in running gzip was less than 5\% of the
checkpoint and restart times.

For DMTCP, all checkpoints are to a local disk (local to the given
computer node), except as noted.  Open~MPI/BLCR checkpointing invokes
an additional step to copy all checkpoint images to a single, central node.

\paragraph{Experimental Configuration.}

Two clusters were employed for the experiments described here.
Sections~\ref{sec:comparison} and~\ref{sec:upc} refer to
a cluster at the Center for
Computational Research at the University of Buffalo. It uses SLURM as
its resource manager, and a common NFS-mounted filesystem.
Each node is equipped with either a Mellanox or QLogic (now Intel)
HCA, although a given partition under which an experiment was run was
always homogeneous (either all Mellanox or all QLogic). The operating
system is RedHat Enterprise Linux~6.1 with Linux kernel version~2.6.32.

Experiments were run using one core per computer.  Hence, the MPI rank
was equal to the number of computers, and so
MPI processes were on separate computer nodes.
Since the memory per node was not fixed,
we set the memory limit per CPU to 3~GB.
Each CPU has a clock rate ranging from 2.13~GHz to 2.40~GHz, and
the specific CPUs can vary.

In terms of software, we used Open~MPI~1.6, DMTCP~2.1 (or a pre-release
version in some cases), Berkeley UPC~2.16.2, and BLCR~0.8.3, respectively.
Open~MPI was run in its default mode, which typically used the RDMA
model for InfiniBand, rather than the send-receive model.
Although, DMTCP version~2.1 was used, the plugin included some additional
bug fixes appearing after that DMTCP release.
For the applications, we used
Berkeley UPC (Unified Parallel~C) version~2.18.0.
The NAS Parallel Benchmark was version~3.1.

Tests of
BLCR under Open~MPI were run by using the Open~MPI checkpoint-restart
service~\cite{HurseyEtAl09}.  Tests of DMTCP for Open~MPI did not use the
checkpoint-restart service.  Under both DMTCP and BLCR, each node
produced a checkpoint image file in a local, per-node scratch partition.
However, the time reported for BLCR includes the time to copy all
checkpoint images to a single node, preventing a direct comparison.

In the case of BLCR, the current Open~MPI checkpoint-restart service
copies each local checkpoint-image to a central coordinator process.
Unfortunately, this serializes part of the parallel checkpoint.  Hence,
checkpoint times for BLCR are not directly comparable to those for DMTCP.

In the case of DMTCP, all tests were run using the
DMTCP default parameters, which include dynamic gzip compression.
Gzip compression results in little compression for numerical data,
and increases the checkpoint time several fold.  Further, DMTCP
is checkpointing to a local ``tmp'' directory on each node.
BLCR checkpoints to a local ``tmp'' directory, and then the
Open MPI service copies them to a central node that saves
them in a different ``tmp'' directory.

\subsection{Scalability of InfiniBand Plugin}
\label{sec:scalability}

Scalability tests for the DMTCP plugin are presented, using
Open~MPI as the vehicle for these tests.
All tests in this section were run at the Massachusetts Green
High-Performance Computing Center (MGHPCC).
Intel Xeon E5-2650 CPUs running at 2~GHz.  Each node is dual-CPU,
for a total of 16~cores per node.  It employs Mellanox HCA adapters.
In addition to the front-end InfiniBand network, there is a Lustre
back-end network.  All checkpoints are written to a local disk,
except for a comparison with Lustre, described later in Table~\ref{tbl:lustre}.

Table~\ref{tbl:scalability} presents a study of scalability for the
InfiniBand plugin.  The NAS MPI test for LU is mmployed.
For a given number of processes, each of classes C, D, and E are
tested provided that the running time for the test is of reasonable length.
The overhead when using DMTCP is analyzed further in Table~\ref{tbl:overhead}.

\begin{table}[ht]
\centering
\begin{tabular}{|l|r|r|r|}
   \hline
NAS        & \hbox{Num.} of~ & Runtime (s) & Runtime (s)~~  \\
benchmark  & processes       & (natively)~~~ & (w/ DMTCP) \\
   \hline
   \hline
LU.C    & 64        & 18.5    & 21.7             \\
   \hline
LU.C    & 128       & 11.5    & 16.1             \\
   \hline
LU.C    & 256       & 7.7     & 12.8             \\
   \hline
LU.C    & 512       & 6.6     & 11.9             \\
   \hline
LU.C    & 1024      & 6.2     & 13.0             \\
   \hline
   \hline
LU.D    & 64        & 292.6   & 298.0            \\
   \hline
LU.D    & 128       & 154.9   & 161.6            \\
   \hline
LU.D    & 256       & 89.0    & 94.8             \\
   \hline
LU.D    & 512       & 53.2    & 61.3             \\
   \hline
LU.D    & 1024      & 30.5    & 39.6             \\
   \hline
LU.D    & 2048      & 26.9    & 40.3             \\
   \hline
   \hline
LU.E    & 512       & 677.2   & 691.6            \\
   \hline
LU.E    & 1024      & 351.6   & 364.9            \\
   \hline
LU.E    & 2048      & 239.3   & 256.4            \\
   \hline
\end{tabular}
\caption{\label{tbl:scalability} Demonstration of scalability:
    running times without DMTCP (natively) and with DMTCP}
\end{table}

In Table~\ref{tbl:overhead}, the overhead derived from
Table~\ref{tbl:scalability} is decomposed into two components:  startup
overhead and runtime overhead.  Given a NAS parallel benchmark,
the total overhead is the difference of the runtime with DMTCP and
the the native runtime (without DMTCP).  Consider a fixed number of
processes on which two different classes of the same benchmark are run.
For example, given the native runtimes
for
two different classes of the LU benchmark (e.g., $t_1$ for LU.C and~$t_2$
for LU.D), and the total overhead in each case ($o_1$ and~$o_2$),
one can derive an assumed startup overhead~$s$ in seconds and
runtime overhead ratio~$r$, based on the formulas:
\begin{align*}
	&o_1 = s + r n_1  \\
	&o_2 = s + r n_2.
\end{align*}

\begin{table}[ht]
\centering
\begin{tabular}{|r|l|r|r|}
\hline
\# processes~ & NAS     & Startup~~~~~ & Slope (runtime          \\
(running LU)           & classes & overhead (s) & overhead in \%)         \\ \hline
64        & C, D       & 3.1          & 0.8                     \\ \hline
128       & C, D       & 4.4          & 1.5                     \\ \hline
256       & C, D       & 5.0          & 0.9                     \\ \hline
512       & D, E       & 7.6          & 1.0                     \\ \hline
1024      & D, E       & 8.7          & 1.3                     \\ \hline
2048      & D, E       & 12.9         & 1.7                     \\ \hline
\end{tabular}
\caption{\label{tbl:overhead} Analysis of Table~\ref{tbl:scalability}
  showing derived breakdown of DMTCP overhead into startup overhead
  and runtime overhead.  (See analysis in text.)}
\end{table}

Table~\ref{tbl:overhead} reports the derived startup overhead
and runtime overhead using the formula above.  In cases where three classes
of the NAS LU benchmark were run for the same number of nodes, the
largest two classes were chosen for analysis.  This decision was made
to ensure that any timing perturbations in the experiment would be
a small percentage of the native runtimes.

The runtime overhead shown in Table~\ref{tbl:overhead} remains in a narrow
range of 0.8\% to 1.7\%.  The startup overhead grows as the cube root
of the number of MPI~processes.

Table~\ref{tbl:ckpt-analysis} below shows the effects on checkpoint
time and checkpoint image size under several configurations.
Note that the first three tests hold constant the number of MPI
processes at~512.  In this situation, the checkpoint size remains
constant (to within the natural variability of repeated runs).
Further, in all cases, the checkpoint time is roughly proportional
to the total size of the checkpoint images on a single node.
A checkpoint time of between 20~MB/s and 27~MB/s was achieved
in writing to local disk, with the faster times occurring
for 16~processes per node (on 16~core nodes).

\begin{table}[ht]
\centering
\begin{tabular}{|l|l|r|r|}
\hline
NAS     & Number of        & Ckpt~~~   & Ckpt~~~~~~  \\
benchmark & processes      & time (s)  & size (MB) \\ \hline\hline
LU.E    & 128$\times$4     & 70.8      & 350       \\ \hline
LU.E    & 64$\times$8      & 136.6     & 356       \\ \hline
LU.E    & 32$\times$16     & 222.6     & 355       \\ \hline\hline
LU.E    & 128$\times$16    & 70.2      & 117       \\ \hline
\end{tabular}
\caption{\label{tbl:ckpt-analysis} Checkpoint times and image sizes
    for the same NAS benchmark, under different configurations.
    The checkpoint image size is for a single MPI process.}
\end{table}

Next, a test was run to compare checkpoint times when using the
Lustre back-end storage versus the default checkpoint to a local disk.
As expected, Lustre was faster.  Specifically, Table~\ref{tbl:lustre} shows
that checkpoint times were 6.5~times faster with Lustre, although restart
times were essentially unchanged.  Small differences in checkpoint image
sizes and checkpoint times are part of normal variation between runs,
and was always limited to less than~5\%.

\begin{table}[ht]
\centering
\begin{tabular}{|l|r|r|r|}
\hline
Disk type  &  Ckpt~~~~~~   & Ckpt~~~~ & Restart~~~   \\
           &  size (MB) & time (s) & time (MB) \\ \hline
local disk &  356       & 232.3    & 11.1      \\ \hline
Lustre     &  365       & 35.7     & 10.9      \\ \hline
\end{tabular}
\caption{\label{tbl:lustre} Comparison with checkpoints to local disk
  or Lustre back-end.  Each case was run for NAS~LU (class~E), and 
  512~processes (32~nodes $\times$ 16~cores per node).
  Each node was writing approximately $16\times 360 = 5.76$~GB
  of checkpoint images.}
\end{table}

Finally, a test was run in which DMTCP was configured not to use its
default gzip compression.  Table~\ref{tbl:gzip}, below, shows that this makes
little difference both for the checkpoint image size and the checkpoint
time.  The checkpoint time is about 5\% faster when gzip is not invoked.

\begin{table}[ht]
\centering
\begin{tabular}{|l|r|r|r|}
\hline
Program and &  Ckpt~~~   & Ckpt~~~~ & Restart   \\
processes   &  size (MB) & time (s) & time (MB) \\ \hline
with gzip &  117       & 70.2       & 23.5      \\ \hline
w/o gzip  &  116       & 67.3       & 23.2      \\ \hline
\end{tabular}
\caption{\label{tbl:gzip} Comparison of checkpointing with and without
  the use of gzip for on-the-fly compression by DMTCP.}
\end{table}

\begin{table*}[bp]
\centering
  \scriptsize
  \begin{tabular}{|l|r|r|r|r|r|r|r|}
   \cline{1-8}
     NAS       & Number of & Runtime (s) & Runtime (s)~~ & Runtime (s) & DMTCP & BLCR~~~ & DMTCP~~~ \\
     benchmark & processes~ & natively~~~~~ & with DMTCP & with BLCR~
                                        & Ckpt (s) & Ckpt (s) & Restart (s) \\
   \cline{1-8}
          & 8 & 224.7 & 229.0 & 240.9 & 7.6 & 16.8 & 2.3 \\
   \cline{2-8}
     LU C & 16 & 116.0 & 117.5 & 118.7 & 5.2 & 16.8 & 2.3 \\
   \cline{2-8}
          & 32 & 61.0 & 64.2 & 64.8 & 3.8 & 16.2 & 2.1 \\
   \cline{2-8}
          & 64 & 32.3 & 35.4 & 34.0 & 2.6 & 20.6 & 2.1 \\

   \cline{1-8}
          & 8 & 885.3 & 886.2 & 887.9 & 1.2 & 3.1 & 0.8 \\
   \cline{2-8}
     EP D & 16 & 442.3 & 447.2 & 448.3 & 1.3 & 3.4 & 1.2 \\
   \cline{2-8}
          & 32 & 223.2 & 225.4 & 227.6 & 1.4 & 4.7 & 3.3 \\
   \cline{2-8}
          & 64 & 115.9 & 118.2 & 122.0 & 1.6 & 8.2 & 1.8 \\
    \hline

   \cline{1-8}
          & 9 & 224.3 & 227.9 & 227.4 & 13.3 & 26.9 & 3.9 \\
   \cline{2-8}
          & 16 & 137.8 & 138.4 & 137.8 & 9.1 & 24.2 & 4.0 \\
   \cline{2-8}
     BT C & 25 & 79.3 & 79.7 & 81.2 & 6.4 & 25.5 & 3.6 \\
   \cline{2-8}
          & 36 & 57.3 & 58.7 & 59.1 & 5.4 & 29.2 & 2.2 \\
   \cline{2-8}
          & 64 & 31.3 & 32.3 & 33.6 & 3.9 & 33.8 & 2.3 \\

   \cline{1-8}
          & 9 & 234.5 & 238.3 & 238.0 & 10.3 & 23.6 & 4.0 \\
   \cline{2-8}
          & 16 & 132.5 & 133.1 & 133.3 & 6.8 & 21.1 & 3.7 \\
   \cline{2-8}
     SP C & 25 & 77.8 & 80.1 & 79.0 & 5.8 & 22.4 & 1.9 \\
   \cline{2-8}
          & 36 & 55.7 & 57.3 & 58.7 & 4.8 & 25.8 & 2.0 \\
   \cline{2-8}
          & 64 & 33.4 & 33.7 & 31.1 & 3.1 & 34.1 & 2.2 \\
   \hline
  \end{tabular}
\caption{\label{tbl:comparison} Comparison of running natively or
 with DMTCP or BLCR.
 Runtime overheads and checkpoint-restart times
 are shown.  (However, BLCR does not report statistics for restart times.)
 Note that for the BT and SP tests, they require
 a number of nodes that is a square of an integer.}
\end{table*}

\subsection{Comparison between DMTCP and BLCR}
\label{sec:comparison}

Table~\ref{tbl:comparison}, below, shows the overall performance of
DMTCP and BLCR.  We chose Open MPI and the NAS Parallel Benchmarks
for a test of performance across a broad test suite.  For the sake of
comparability with previous tests on the checkpoint-restart service of
Open MPI~\cite{HurseyEtAl09}, we include the previously used NAS tests:
LU~C, EP~D, BT~C and SP~C.  An analysis of the performance must consider
both runtime overhead and times for checkpoint and restart.

\paragraph{Runtime overhead.}

The runtime overhead is the overhead for running a program under a
checkpoint-restart package as compared to running natively.  No checkpoints
are taken when measuring runtime overhead.  Table~\ref{tbl:comparison}
shows that neither DMTCP nor BLCR has significant runtime overhead.
For longer running program the runtime overhead is in the range of 1\% to~2\%.
A typical runtime overhead is in the range of 1 to 5~seconds, with 3~seconds
being common.  Since these overhead times do not correlate with the
length of time during which the program was run, we posit that they
reflect a constant overhead that is incurred primarily at the time of
for program startup.

\paragraph{Checkpoint/Restart times.}

Table~\ref{tbl:comparison} also shows checkpoint and restart times.
These times are as reported by a central DMTCP coordinator, or an
Open~MPI coordinator in the case of BLCR.

Note that the restart time for DMTCP was
typically under 4~seconds, even for larger computations using 64~computer
nodes.  The BLCR package did not report restart times.

Checkpoint times are particularly important for issues of fault tolerance,
since checkpoints are by far the more common operation.  Surprisingly,
the checkpoint time for DMTCP falls with an increasing number of nodes
(except in the case of EP: {\em Embarrassingly Parallel}), while the
checkpoint time for BLCR increases slowly with an increasing number of nodes.

For DMTCP, checkpoint times are smaller with an increasing number of nodes
because the NAS benchmarks leave fixed the total amount of work.
Thus, twice as many nodes implies half as much work.  In the case
of most of the NAS benchmarks, the reduced work per node seems to imply
a reduced checkpoint image size (except for EP: {\em Embarrassingly
Parallel}).  This accounts for the faster checkpoint times as the
number of nodes increases.

In contrast, the checkpoint-restart service of Open MPI, the times for
checkpointing with Open~MPI/BLCR are often roughly constant.  We estimate
this time is dominated by the last phase, in which Open~MPI copies the
local checkpoint images to a single, central node.

\subsection{Checkpointing under UPC: a non-MPI Case Study}
\label{sec:upc}

In our study of Berkeley UPC, is
based on a port of the
NAS parallel benchmarks at George Washington University~\cite{upc-nas}.
Since that did not include a port of the LU benchmark, we switch to
considering FT in this example.

We compiled the Berkeley UPC package to run natively over InfiniBand,
so that it did not use MPI.  We chose to test on the FT~B NAS benchmark,
as ported to run on UPC.
The FT benchmark was used because that port does not support the
more communication-intensive LU benchmark.
Table~\ref{tbl:upc} shows that the native runtimes for FT~B under UPC
are compared to the time for~MPI in Table~\ref{tbl:comparison}.
DMTCP total run-time overhead
ranges from 4\% down to less than 1\%.  We posit that the higher
overhead of 4\% is due to the extremely short running time in the case
of 16~processes, and is explained by significant startup overhead,
consistent with Table~\ref{tbl:overhead}.

Note that BLCR could not be tested in this regime, since BLCR depends
on the Open~MPI checkpoint-restart service for its use in distributed
computations.

\begin{table}[htp]
\centering
  \begin{tabular}{|r|r|r|r|r|}
   \cline{1-5}
     Number of & Runtime~~~   & Runtime w/ & Ckpt & Restart \\
     processes~ & natively (s) & DMTCP (s)  & (s)~~& (s)~~   \\
   \hline
   4 & 123.5 & 124.2 & 27.6 & 9.7 \\
   \hline
   8 & 64.2 & 65.1 & 21.9 & 8.9 \\
   \hline
   16 & 34.2 & 35.5 & 16.3 & 7.0 \\
   \hline
  \end{tabular}
\caption{\label{tbl:upc} Runtime overhead and Checkpoint-restart times
for UPC FT~B running under DMTCP}
\end{table}

\subsection{Migrating from InfiniBand to TCP sockets}
\label{sec:ib2tcp}
Some traditional checkpoint-restart services, such as that for
Open~MPI~\cite{HurseyEtAl09}, offer the ability to checkpoint
over one network, and restart on a second network.  This is
especially useful for interactive debugging.  A set of checkpoint
images from an InfiniBand-based production cluster can be copied
to an Ethernet/TCP-based debug cluster.  Thus, if a bug is
encountered after running for hours on the production cluster,
the most recent checkpoints can be used to restart on the debug
cluster under a symbolic debugger, such as GDB.  The approach
of this paper offers the added feature that the Linux kernel
on the debug cluster does not have to be the same as on the
production cluster.

\subsubsection{IB2TCP: Ping-pong}
We tested the IB2TCP plugin with a communication intensive ping-pong
example InfiniBand program from the OFED distribution.
In this case, a smaller development cluster was used,
with 6-core Xeon~X5650 CPUs and a Mellanox HCA for InfiniBand.
Gigabit Ethernet was used for the Ethernet portion.
Parameters were set to run over 100,000 iterations.

\begin{table}[htp]
\centering
  \begin{tabular}{|l|r|r|}
   \cline{1-3}
     ~~~~~~~~Environment & Transfer & Transfer rate \\
                         & time (s) & (Gigabits/s)~ \\
   \hline
    IB (w/o DMTCP) & 0.9 & 7.2\\
   \hline
    DMTCP/IB (w/o IB2TCP) & 1.2 &  5.7\\
   \hline
    DMTCP/IB2TCP/IB & 1.4 &  4.6\\
   \hline
    DMTCP/IB2TCP/Ethernet & 65.7 & 0.1 \\
   \hline
  \end{tabular}
\caption{\label{tbl:ib2tcp} Transfer time variations when using two nodes
	on InfiniBand
	versus Gigabit Ethernet hardware, as affected by the DMTCP InfiniBand
	plugin and the IB2TCP plugin; 100,000 iterations of ping-pong,
	for a total transfer size of 819~MB}
\end{table}

Table~\ref{tbl:ib2tcp} presents the results.
These results represent a worst case, since a typical MPI program is not
as communication-intensive as the ping-ping test program.  The times
with IB2TCP were further degraded due to the current implementation's use
of an in-memory copy.  We hypothesize that the transfer rate under Gigabit
Ethernet was also limited by the speed of the kernel implementation.

\subsubsection{IB2TCP: NAS LU.A.2 Benchmark}
Next, a preliminary test of IB2TCP for Open~MPI, on only two nodes,
is presented.
This test was conducted on the the MGHPCC cluster, as described
in Section~\ref{sec:scalability}.

Table~\ref{tbl:lu-ib2tcp} shows various times for NAS LU.A.2 benchmark for
migrating from InfiniBand to Ethernet using the
IB2TCP plugin. As can be seen, the InfiniBand and IB2TCP plugin do not add any
considerable overhead to the run time of the application. However, when the
process is migrated from InfiniBand to Ethernet, the runtime increases drastically.
A runtime overhead of 67\% is seen when the computation is restarted on two
nodes. The runtime overhead further increases to 142\% when the entire
computation is restarted on a single node.
\begin{table}[htp]
\centering
  \begin{tabular}{|l|r|}
   \hline
     ~~~~~~~~Environment     & RunTime (s) \\
   \hline
    IB (w/o DMTCP)           & 26.61 \\
   \hline
    DMTCP/IB (w/o IB2TCP)    & 27.81 \\
   \hline
    DMTCP/IB2TCP/IB          & 27.38 \\
   \hline
    DMTCP/IB2TCP/Ethernet    & 45.75 \\
    (restart on two nodes)   &       \\
   \hline
    DMTCP/IB2TCP/Ethernet    & 66.34 \\
    (restart on a single node) &       \\
   \hline
  \end{tabular}
\caption{\label{tbl:lu-ib2tcp}
Runtime variations for LU.A.2 benchmark when using two nodes on InfiniBand
versus Gigabit Ethernet hardware, as affected by the DMTCP InfiniBand plugin
and the IB2TCP plugin.  The runtime does not involve the checkpoint and restart
times.}
\end{table}

\section{Future Work}
\label{sec:futureWork}

The current work supports only a homogeneous InfiniBand architecture.
For example, one cannot checkpoint on a node using an Intel/Qlogic
HCA and restart on a different node that uses a Mellanox HCA.  This is
because the checkpoint image already includes a low-level library to
support a particular HCA.  Future work will extend this implementation
to support heterogeneous InfiniBand architectures by re-loading the
low-level library, as described in Section~\ref{sec:limitations}.

Further, the experimental timings reported here did not employ any particular
tuning techniques.  There are opportunities to reduce the run-time
overhead by reducing the copying of buffers.

\paragraph{Avoiding conflict of virtual ids after restart:}
In typical MPI implementations memory region keys (rkey), queue pair
numbers (qp\_num), and local ids (lid) are all exchanged out-of-band.
Since virtualized ids are passed to the target application, it is
the virtualized ids that are passed out-of-band.  The remote plugin
is then responsible for translating the virtual ids to the real ids
known to the InfiniBand hardware, on succeeding InfiniBand calls.

The current implementation ensures that this is possible, and that there
are no conflicts prior to the first checkpoint, as described in
Section~\ref{sec:virtualization}.  In typical InfiniBand applications,
queue pairs are created only during startup, and so
all rkeys, qp\_nums and lids will be assigned prior to the first
checkpoint.  However, it is theoretically possible for an application
to create a new queue pair, memory region, or to query its local id
after the first restart.

The current implementation assigns the virtual~id
to be the same as the real~id at the time of the initial creation
of the InfiniBand object.  (After restart, the InfiniBand hardware may
assign a different real~id, but the virtual~id for that object
will remain the same.)  If an object is created after restart, the
real~id assigned by InfiniBand may be the same as for an object
created prior to checkpoint.  This would create a conflict of the
corresponding virtual~ids.

Two solutions to this problem are possible.  The simplest is to use
DMTCP's publish-subscribe feature to generate globally unique virtual
rkeys, and update a global table of virtual-to-real rkeys.  In particular,
one could use the existing implementation before the first checkpoint,
and then switch to a publish-subscribe implementation after restart.
A second solution is to choose the virtual rkeys in a globally unique
manner, similarly to the globally unique protection domain ids of the
current plugin.

\section{Conclusion}
\label{sec:conclusion}

A new approach to distributed transparent checkpoint-restart over
InfiniBand has been demonstrated.  This direct approach accommodates
computations both for MPI and for UPC (Unified Parallel~C).
The approach uses a mechanism similar to that of a shadow device
driver~\cite{SwiftEtAl04,SwiftEtAl06}.  In tests on the NAS LU parallel
benchmark, a run-time overhead of between 0.7\% and 1.9\% is found on a
computation with up to 2,048 MPI processes.  Startup overhead is up to
13~seconds, and grows as the cube root of the number of MPI processes.
Checkpoint times are roughly proportional to the total size of all
checkpoint images on a single computer node.  In one example, checkpoint
times varied by a factor of~6.5 (from 232~seconds to 36~seconds), depending
on whether checkpoint images were written to a local disk or to a faster,
Lustre-based back-end.  In each case, 16~MPI processes per node (512
processes in all) wrote a total of approximately 5.8~GB per node.

Further, the new approach provides a viable checkpoint-restart mechanism
for running UPC natively over InfiniBand --- something that previously did
not exist.  finally, an IB2TCP plugin was demonstrated that allows users
to checkpoint a computation on a production cluster under InfiniBand,
and to restart on a smaller debug cluster under Ethernet --- in a manner
suitable for attaching an interactive debugger.  This mode is analogous to
the interconnection-agnostic feature of the Open~MPI checkpoint restart
service~\cite{HurseyEtAl09}, but it has an added benefit in that the
production cluster and the debug cluster do not have to have the same
Linux kernel image.

\section*{Acknowledgment}

We are grateful to facilities provided at several institutions with
which to test over a variety of configurations.  We would like
to thank:
\hbox{L.} Shawn Matott (\hbox{U.}~of Buffalo, development and
	 benchmarking facilities);
Henry Neeman (Oklahoma University, development facilities);
Larry Owen and Anthony Skjellum (the University of Alabama at
Birmingham, facilities based on NSF grant CNS-1337747);
and
the Massachusetts Green High Performance Computing Center (facilities
for scalability testing).
Dotan Barak provided helpful advice on the implementation of OpenFabrics
InfiniBand.  Jeffrey \hbox{M.}~Squyres and Joshua Hursey provided helpful
advice on the interaction of Open~MPI and InfiniBand.  Artem Polyakov
provided advice on using the DMTCP batch-queue (resource manager) plugin.

\bibliographystyle{abbrv}

\end{document}